\documentclass[sigconf,table,natbib=true,anonymous=false]{acmart}
\renewcommand\footnotetextcopyrightpermission[1]{} 
\usepackage[normalem]{ulem}
\useunder{\uline}{\ul}{}
\usepackage{multirow}
\usepackage{makecell}
\usepackage{adjustbox}
\usepackage{enumitem}
\usepackage{tabularx}
\usepackage{subfigure}
\usepackage{tikz}

\AtBeginDocument{%
  \providecommand\BibTeX{{%
    \normalfont B\kern-0.5em{\scshape i\kern-0.25em b}\kern-0.8em\TeX}}}

\usepackage{mathtools}

\setcopyright{acmcopyright}
\copyrightyear{2018}
\acmYear{2018}
\acmDOI{XXXXXXX.XXXXXXX}

\acmConference[Conference acronym 'XX]{Make sure to enter the correct
  conference title from your rights confirmation emai}{June 03--05,
  2018}{Woodstock, NY}
\acmPrice{15.00}
\acmISBN{978-1-4503-XXXX-X/18/06}

\begin{document}

\title{Unsupervised Large Language Model Alignment for Information Retrieval via Contrastive Feedback}


\author{Qian Dong}
\email{dq22@mails.tsinghua.edu.cn}
\affiliation{%
  \institution{Quan Cheng Laboratory\&}
  \institution{DCST, Tsinghua University\&}
  \institution{Zhongguancun Laboratory}
  \city{Beijing}
  \country{China}}
\author{Yiding Liu}
\email{liuyiding.tanh@gmail.com}
\affiliation{%
  \institution{Baidu Inc.}
  \city{Beijing}
  \country{China}}
\author{Qingyao Ai}\authornote{Corresponding author.}
\email{aiqy@tsinghua.edu.cn}
\affiliation{%
  \institution{Quan Cheng Laboratory\&}
  \institution{DCST, Tsinghua University\&}
  \institution{Zhongguancun Laboratory}
  \city{Beijing}
  \country{China}}
\author{Zhijing Wu}
\email{zhijingwu@bit.edu.cn}
\affiliation{%
  \institution{School of Computer Science and Technology}
  \institution{Beijing Institute of Technology}
  \city{Beijing}
  \country{China}}
\author{Haitao Li}
\email{liht22@mails.tsinghua.edu.cn}
\affiliation{%
  \institution{Quan Cheng Laboratory\&}
  \institution{DCST, Tsinghua University\&}
  \institution{Zhongguancun Laboratory}
  \city{Beijing}
  \country{China}}
\author{Yiqun Liu}
\email{yiqunliu@tsinghua.edu.cn}
\affiliation{%
  \institution{Quan Cheng Laboratory\&}
  \institution{DCST, Tsinghua University\&}
  \institution{Zhongguancun Laboratory}
  \city{Beijing}
  \country{China}}
\author{Shuaiqiang Wang}
\email{shqiang.wang@gmail.com}
\affiliation{%
  \institution{Baidu Inc.}
  \city{Beijing}
  \country{China}}
\author{Dawei Yin}  
\email{yindawei@acm.org}
\affiliation{%
  \institution{Baidu Inc.}
  \city{Beijing}
  \country{China}}
\author{Shaoping Ma}
\email{msp@tsinghua.edu.cn}
\affiliation{%
  \institution{Quan Cheng Laboratory\&}
  \institution{DCST, Tsinghua University\&}
  \institution{Zhongguancun Laboratory}
  \city{Beijing}
  \country{China}}
\renewcommand{\shortauthors}{Qian Dong et al.}
\begin{abstract}
Large language models (LLMs) have demonstrated remarkable capabilities across various research domains, including the field of Information Retrieval (IR). 
However, the responses generated by off-the-shelf LLMs tend to be generic, i.e., cannot capture the \textbf{distinctiveness} of each document with similar content. 
This limits the performance of LLMs in IR because finding and distinguishing relevant documents from substantial similar documents is a typical problem in many IR tasks.
To address this issue, we propose an unsupervised alignment method, namely \textit{Reinforcement} \textit{Learning} from \textit{Contrastive} \textit{Feedback} (\textbf{RLCF}), empowering LLMs to generate both high-quality and context-specific responses.
Our approach constructs unsupervised contrastive feedback signals based on similar document groups, and adopts a reward function, named group-wise reciprocal rank, to optimize LLMs within a standard Proximal Policy Optimization. 
We conduct extensive experiments to evaluate the effectiveness of RLCF on LLMs built with different languages and parameter sizes on multiple downstream IR applications.
RLCF significantly outperforms existing alignment methods, and RLCF-optimized LLMs demonstrate considerable improvement in generating responses with distinctiveness.
\end{abstract}




%


\maketitle
\section{Introduction}
Large Language Models (LLMs) have demonstrated promising performances across a wide range of research fields, including information retrieval (IR).
IR aims to fulfil information needs of individuals though locating relevant documents from large-scale corpus~\cite{cheng2023layout}, which plays a fundamental role in the digital era~\cite{dong2022disentangled,chen2022axiomatically,yang2022enhance}.
Previous studies have utilized LLMs in IR tasks such as generating search snippets~\cite{Reid_2023, liu2023evaluating, huang2023citation, kou2023automated}, conducting query or document expansion~\cite{gospodinov2023doc2query, ma2023pre, wang2023query2doc}, creating training data for retrieval models~\cite{wang2021gpl, bonifacio2022inpars,dai2022promptagator}, and etc. These studies have demonstrated the significant potential of LLMs in IR. 

\begin{figure}
    \centering
    \includegraphics[width=\linewidth]{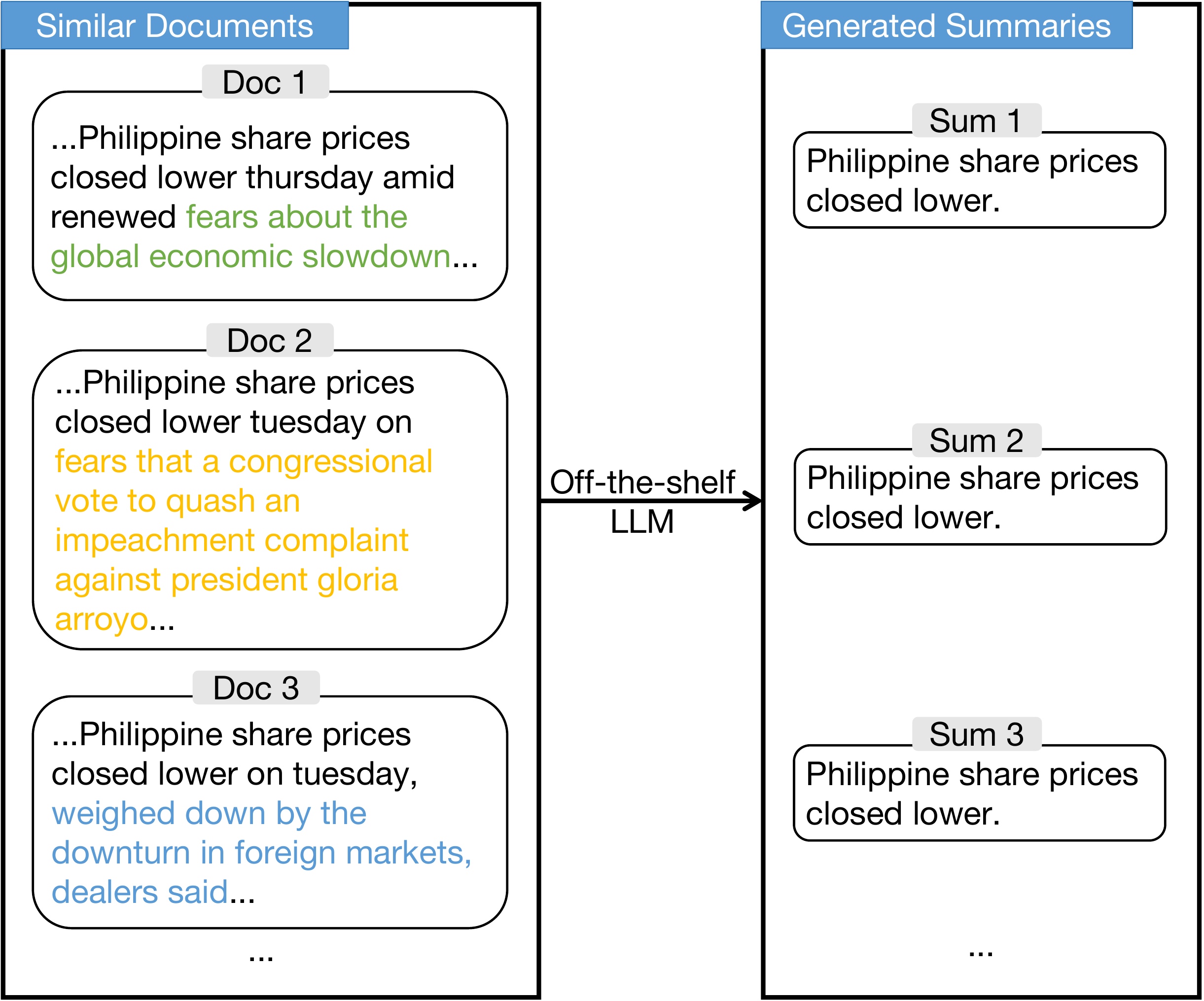}
    \caption{Illustrations of LLMs application in document summarization for similar documents. The distinctive parts of each document are highlighted in different colors.}
    \label{fig:llminir}
    \vspace{-4mm}
\end{figure}

However, while widely adopted, applying an off-the-shelf LLM directly to IR tasks could be suboptimal. 
One of key reasons is the misalignment between the capabilities of off-the-shelf LLMs and the needs of IR tasks, particularly the capability to provide responses that capture the \textbf{distinctiveness} of each document among similar documents~\cite{dai2022promptagator, gospodinov2023doc2query}. 
For instance, Figure~\ref{fig:llminir} show the results of FLAN-T5~\cite{chung2022scaling} when used for document summarization, a representative IR task. 
As we can see, off-the-shelf FLAN-T5 generates the same summary for three different documents, making them indistinguishable to human.
Using such summaries as document representations or search snippets is not acceptable in search engines, as they fail to assist users in quickly and accurately locating their desired documents.

To empower LLMs with desired response attributes, one of the most widely adopted methods is to conduct LLM alignment. 
The idea of LLM alignment is to construct feedback to LLM's responses based on the preferred attributes (e.g., helpfulness~\cite{ouyang2022training}, harmlessness~\cite{lee2023rlaif, yang2023rlcd, dai2023safe}, etc.) and use them to optimize LLMs. 
Depending on how the feedback is collected, existing LLM alignments can be broadly categorized into human-based methods and model-based methods. 
Unfortunately, none of the existing alignment methods can be used to improve distinctiveness of response from LLMs.
\begin{itemize}[left=0.05in]
    \item While human-based methods such as RLHF~\cite{ouyang2022training} are flexible and generic, the cost of collecting large-scale human feedback is expensive and usually prohibitive in IR scenarios. 
    \item While existing model-based methods such as RLAIF~\cite{lee2023rlaif} and RLCD~\cite{yang2023rlcd} do not require human in the loop, they are inherently incapable of capturing distinctiveness, as their feedback signals (i.e., reward scores) are computed in a point-wise manner.
    Specifically, the distinctive information of an LLMs response must be specified in the context of other similar responses, and thus the feedback signals of distinctiveness can only be derived from differentiating multiple responses in a group-wise manner, where existing methods are clearly infeasible.
\end{itemize}

\begin{figure}
    \centering
    \includegraphics[width=0.9\linewidth]{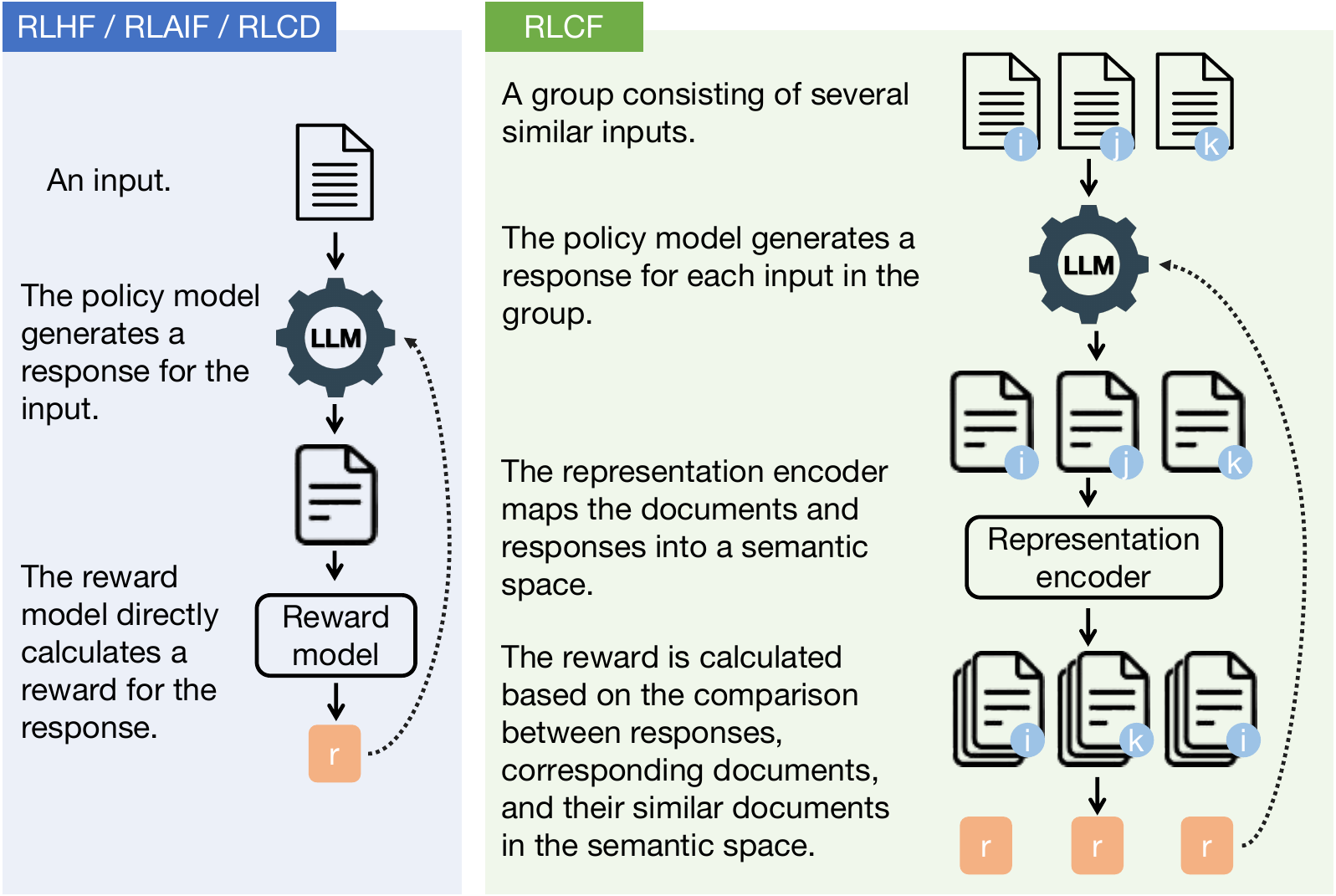}
    \caption{The comparison between existing alignment methods and RLCF. The dotted line represents that the reward score is returned to LLM for PPO optimization.}
    \label{fig:comparison}
    \vspace{-5mm}
\end{figure}

To address these limitations and improve the distinctiveness of LLMs responses, we propose a novel optimization framework called Reinforcement Learning from Contrastive Feedback (\textbf{RLCF}). 
RLCF is a \textbf{group-wise} alignment method that constructs feedback of distinctiveness for a group of similar inputs, and it is fully \textbf{unsupervised}.
The workflow of RLCF is depicted in Figure~\ref{fig:comparison}. Specifically, we first identify a group of similar documents for each document in the corpus. 
Then, each document in the group is fed to an LLM to obtain a corresponding response. For example, the responses could be summaries of long documents for a summarization task.
Next, we construct contrastive feedback from the group of LLM-generated responses, using a reward function named group-wise reciprocal rank.
Finally, the LLM is optimized based on the contrastive feedback with the standard Proximal Policy Optimization (PPO) algorithm~\cite{schulman2017proximal}.

To demonstrate the effectiveness of RLCF in optimizing the distinctiveness of LLMs responses, we conduct comprehensive experiments, which include three representative IR tasks, various scales of LLMs (ranging from 770M to 11B) with both encoder-decoder and decoder-only architectures, and both Chinese and English corpora. 
In particular, we evaluate RLCF on the tasks of document summarization, document expansion and data augmentation. Note that the performances on these tasks are essentially sensitive to the distinctiveness of the LLM-generated responses (e.g., summaries).
For document summarization, we directly evaluate the quality of the LLM-generated summaries, measured by human evaluation and an automatic metric called Rouge-diff.
To assess the effectiveness of RLCF-optimized LLMs in document expansion and data augmentation, we evaluate the LLMs responses through downstream document retrieval tasks, i.e., comparing the performances of retrieval models enhanced by LLMs with different alignment methods. 
The results indicate that, compared to existing alignment methods, RLCF can significantly improve the distinctiveness of LLMs responses, which can largely benefit various IR applications.

We summarize our main contributions as follows:
\begin{itemize}
    \item To the best of our knowledge, we are the first to study the distinctiveness of LLMs response and demonstrate its significance in powering IR applications.
    \item We propose a novel framework namely RLCF which utilizes the contrastive feedback to unsupervised align the capability of LLMs with the needs of IR.
    \item The experimental results demonstrate the effectiveness of RLCF on both English and Chinese LLMs as well as various parameter scales and architectures.
\end{itemize}

\section{Related Work}
\subsection{Large Language Models}
Recently, LLMs are emerged and boost many natural language processing tasks.
The architecture of LLMs, particularly the Transformer~\cite{vaswani2017attention}, leads to significant improvements in capturing textual semantics.
This advancement empowers many influential models such as BERT~\cite{devlin2018bert} and GPT~\cite{radford2018improving}.
These models pave the way for subsequent advancements like GPT-2~\cite{radford2019language} and GPT-3~\cite{brown2020language}, with increasing model sizes and capabilities.
The training pipeline of LLMs also earned significant attention in recent years due to its pivotal role in enabling models like GPT to exhibit remarkable language understanding and generation capabilities. 
Pre-training is a cornerstone of training LLMs and involves training the model on a massive corpus to learn linguistic patterns and structures, leveraging the tasks such as masked language modeling~\cite{devlin2018bert}, next token prediction~\cite{radford2018improving} and etc. 
By utilizing large-scale pre-training, LLMs acquire a general understanding of language, making them available for various downstream tasks.
Supervised Fine-Tuning (SFT) involves training LLMs on task-specific datasets with labeled examples. 
This stage adapts the generic linguistic knowledge acquired during pre-training to specific tasks, such as sentiment analysis~\cite{dong2021latent}, text classification~\cite{dong2021legal,dong2022incorporating}, and dialogues~\cite{ouyang2022training}.
Alignment technique facilitates LLMs in learning from the generated responses and environmental feedback, thereby aligning the capability with the desired attribute. 
The environment feedback could be from human~\cite{ouyang2022training} or other models~\cite{bai2022constitutional}.
This approach has shown promise in improving the helpfulness and harmlessness of LLMs.


\subsection{Alignment for LLM}
Alignment techniques, which aim to ensure that language models act in accordance with human values or desired attributes, have garnered significant research attention.
This surge of interest is primarily attributed to the widespread proliferation and increasing impact of language models in recent years. 
In recent years, researchers have increasingly turned their focus to leveraging human feedback as a valuable resource for optimizing language models~\cite{bohm2019better,ziegler2019fine,stiennon2020learning}.
Reinforcement learning is typically employed for such optimization, leading to the development of a class of methods referred to as RLHF (Reinforcement Learning from Human Feedback). 
RLHF~\cite{ouyang2022training} leverages human-provided reward signals to guide the training process of language models, enhancing their performance in various natural language generation tasks.
Despite the superior performance, a major drawback of RLHF is the requirement for extensive manual labor to provide feedback, making it expensive and time-consuming. 
Recently, RLAIF~\cite{lee2023rlaif} utilizes the feedback from the LLM itself to train a reward model, and aligns the capability of LLM in a same manner with RLHF.
RLCD~\cite{yang2023rlcd} utilizes the positive and negative prompts to output preferred responses, thereby training the reward model.

\subsection{LLM Applications in IR}
\noindent\textbf{Document Summarization.}
Document summarization is a vital research area in information retrieval.
Here, we provide an overview of document summarization.
Extractive summarization methods~\cite{narayan2018ranking,zhou2018neural,liu2019fine,koh2022empirical} select sentences or phrases directly from the input document to form a summary. 
Abstractive summarization~\cite{rush2015neural,see2017get,yadav2023fine,borah2022comparative} approach imitates human that comprehends a source document and writes a summary based on the salient concepts of the document
Multi-document summarization~\cite{cao2015ranking,zheng2019subtopic,bravzinskas2019unsupervised} concentrates on generating concise summaries from a cluster of topic-related documents. 
Besides, PLMs, such as BART~\cite{lewis2019bart}, GPT-2~\cite{radford2019language}, and T5~\cite{raffel2020exploring}, are also be used for multi-document summarization task~\cite{pang2021agreesum,alambo2020topic,su2020caire}.

\noindent\textbf{Language Model for Document Expansion.}
LLMs are widely used to supplement missing information, thereby mitigating issues associated with data sparsity~\cite{wu2023survey} or information gaps~\cite{zhu2023large}.
The vocabulary mismatch between query and document could be effectively alleviate by document expansion using language models~\cite{nogueira2019doc2query,gospodinov2023doc2query, wang2023query2doc}.
Doc2Query~\cite{nogueira2019document} predicts which queries will be issued for a given document and then expands it with those predictions with a vanilla sequence-to-sequence model, trained using datasets consisting of pairs of query and relevant documents.
DocT5Query~\cite{nogueira2019doc2query}, employing T5~\cite{chung2022scaling} as its backbone in the Doc2Query framework, achieves remarkable performance, illustrating that an enhanced backbone results in superior improvements.
Doc2Query-~\cite{gospodinov2023doc2query} illustrates the significance of query quality in document expansion. The research suggests that eliminating low-quality queries can enhance the effectiveness of Doc2Query.

\noindent\textbf{Language Model for Data Augmentation.}
Data augmentation is an effective strategy to address the challenge of limited training sample sizes.
This challenge is especially prominent in the zero-shot learning scenario, which can be viewed as a cold-start problem.
Owing to LLMs' superior language comprehension capabilities, they are extensively employed for data augmentation in numerous research fields, like text classification~\cite{dai2023chataug}, multilingual commonsense reasoning~\cite{whitehouse2023llm}, dense retrieval~\cite{wang2021gpl,bonifacio2022inpars,izacard2021unsupervised,dai2022promptagator} and etc.
The distinctiveness of responses from LLMs plays a pivotal role in data augmentation, particularly in dense retrieval.
Dense retrieval is trained using query and document pairs, which draws extensive attention from both academia and industry due to its superior performance when applied to the documents it has already been trained on~\cite{dong20233,li2023sailer}.
However, in practical search engines, a large number of new web documents are emerged daily, which often leads to a collapse in the performance of dense retrieval methods with respect to these new documents~\cite{thakur2021beir,dai2022promptagator}.
Therefore, boosting the zero-shot performance of dense retrieval on new documents is a crucial challenge, in which LLMs play a pivotal role~\cite{izacard2021unsupervised,thakur2021beir,dai2022promptagator}.

\section{Reinforcement Learning from Contrastive Feedback}
\label{sec:method}
In this section, we present the details of our proposed framework, Reinforcement Learning from Contrastive Feedback (RLCF).
It facilitate LLMs to capture fine-grained distinctions in similar input documents and output responses that are more distinctive.

\subsection{Motivation}
Existing alignment methods, such as RLHF~\cite{ouyang2022training}, RLAIF~\cite{lee2023rlaif}, RLCD \cite{yang2023rlcd}, etc., have demonstrated effectiveness in adjusting response attributes (e.g., helpfulness).
However, these methods can hardly be utilized to enhance the distinctiveness of LLMs response due to two limitations.
First, the feedback signals of existing alignment methods are not applicable for our task at hand. In particular, human feedback can be carefully-designed to imply distinctiveness, yet is clearly too expensive to scale. On the other hand, model-based feedback has more manageable cost since it relies on a model but a human to generate accurate feedback. Unfortunately, as shown in our experiments, existing LLMs struggle to provide accurate feedback w.r.t. distinctiveness.
Second, the feedback computation in existing alignment methods follows a point-wise input manner.
As shown in the left part of Figure~\ref{fig:comparison}, a reward score is computed for a single input in the existing methods, while overlooking the relationships between inputs. 
Consequently, the subtle distinctions among similar inputs are neglected, which could result in trivial or less informative outputs.
Therefore, a natural question is: \textbf{How to construct a group-wise feedback with high-quality and low-cost to enhance the distinctiveness of LLMs' responses?}

\begin{figure}
    \centering
    \includegraphics[width=0.85\linewidth]{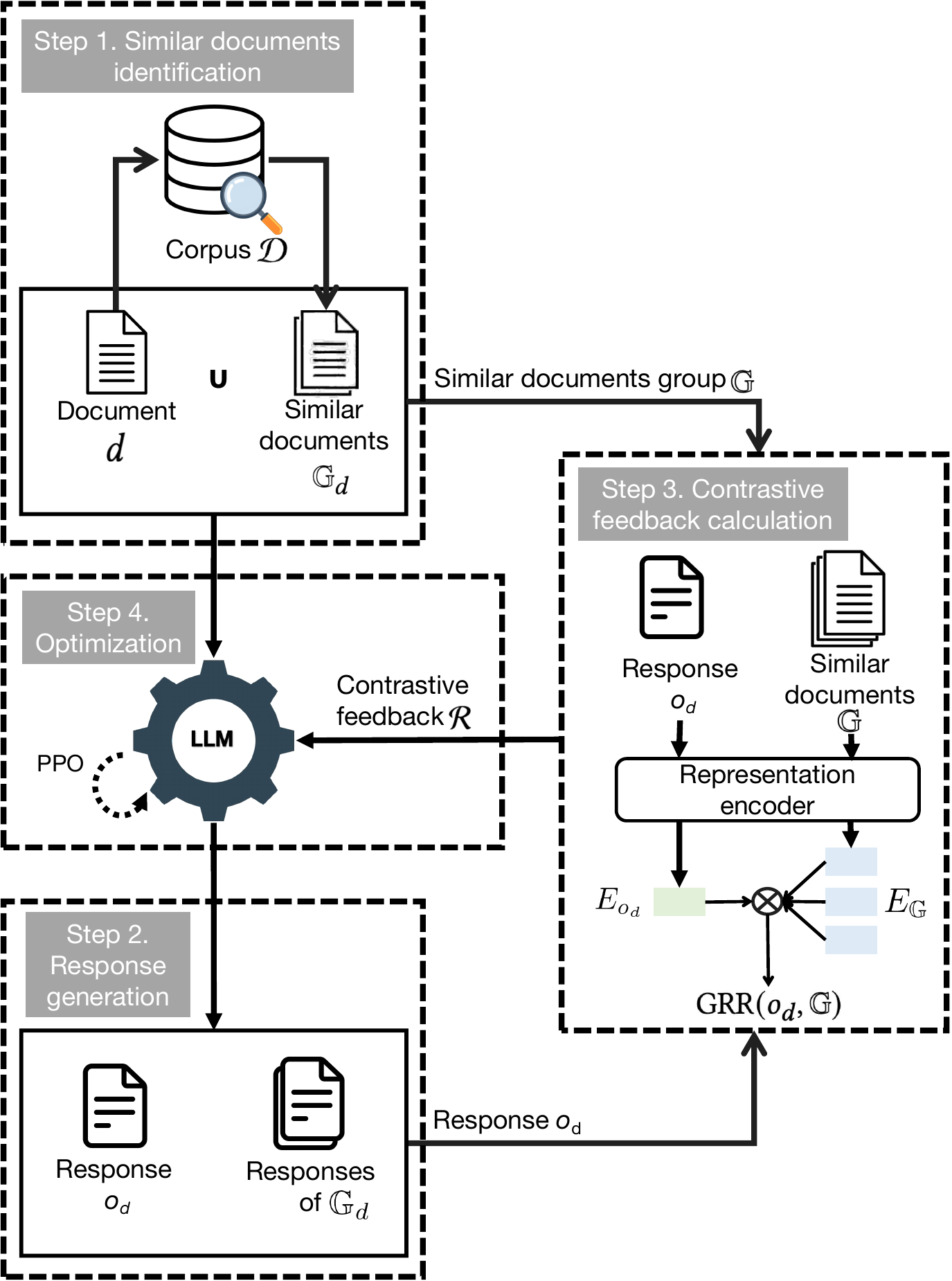}
    \caption{The framework of RLCF. We take the response $o_d$ as an example for illustration of group-wise contrastive feedback calculation. The green and blue rectangles represent the embedding of response and documents, respectively. The $\otimes$ represents the inner production operation between the embedding $E_{o_d}$ and $E_{\mathbb{G}}$.}
    \label{fig:rlcf}
    \vspace{-5mm}
\end{figure}

To answer the question, we propose an unsupervised alignment method, namely \textit{Reinforcement} \textit{Learning} from \textit{Contrastive} \textit{Feedback} (\textbf{RLCF}). As shown in Figure~\ref{fig:rlcf}, 
we first construct data for formulating contrastive feedback, including similar documents identification and response generation.
After that, the rest part in this figure outlines the process of optimizing an LLM with the contrastive feedback, which teaches the LLM to identify more distinctive information from a document.
The commonly used notations are summarized in Table~\ref{table:notation}.


\begin{table}[]
\caption{The notations used in this paper.}
\label{table:notation}
\begin{tabularx}{0.48\textwidth}{l|X}
\hline\hline
Notations   & Descriptions                                          \\ \hline
$q$           & The query used to search a document.                  \\ \hline
$d$           & A document from corpus.                               \\ \hline
$\mathcal{D}$           & The corpus of documents.                              \\ \hline
$\mathbb{G}_d$          & The similar documents of document $d$.            \\ \hline
$o_d$          & The response of LLM for document $d$.                   \\ \hline
$\pi$          & The original parameters of LLM.                       \\ \hline
$\pi_{\phi}^{RL}$   & The optimized RL policy.                              \\ \hline
$\mathcal{R}$           & The reward used to optimize LLM, including the penalty term.                      \\ \hline
GRR & The reward function, referred to as \textbf{g}roup-wise \textbf{r}eciprocal \textbf{r}ank (GRR).                                     \\ \hline
Inst        & An instruction used for response generation. \\ \hline\hline
\end{tabularx}
\vspace{-4mm}
\end{table}




\subsection{Data Construction}
\label{sec:dataConstruction}
\noindent\textbf{Similar Documents Identification.}
To facilitate the capacity of LLMs for capturing subtle distinctions among documents, 
we first need to gather groups of similar documents
for computing contrastive feedback.
To avoid the high cost of data labeling, we leverage an unsupervised dual-encoder to construct each group of similar documents.

In particular, we randomly select a document $d$ in the corpus $\mathcal{D}$, and retrieve its top-K most similar documents to form the similar documents $\mathbb{G}_d$ of document $d$, which can be formally defined as
\begin{equation}
    \label{eq:group}
    \mathbb{G}_{d_i} = \{d_j~|~\underset{\text{top-}\mathrm{K}}{\mathrm{argmax}}\ S\left(d_i, d_j\right), \forall d_j \in \mathcal{D}, i\neq j\},
\end{equation}
where $S(d_i, d_j)$ denotes the similarity score between $d_i$ and $d_j$. 
Taking a standard dual-encoder based dense retriever as an example, the similarity between documents are computed as
\begin{equation}
    \label{eq:similarity}
    S(d_i, d_j)=E_{d_i}\otimes E_{d_j}.
\end{equation}
Here, the $\otimes$ means the inner production operation, and 
\begin{equation}
    \label{eq:encodeT}
    E_d = \mathrm{Avg\_Pooling}\left(\mathrm{M}\left(d\right)\right),
\end{equation}
which is the average pooling of the last layer's token representations produced by the encoder $\mathrm{M}$.

\noindent\textbf{Response Generation.}
Next, for each group of similar documents $\mathbb{G} = \left\{d\cup\mathbb{G}_d\right\}$, we use an LLM to generate a response for each document within the group. 
These response could be a query or the summary w.r.t. the document, or any other desired output.

\begin{figure}
    \centering
    \includegraphics[width=\linewidth]{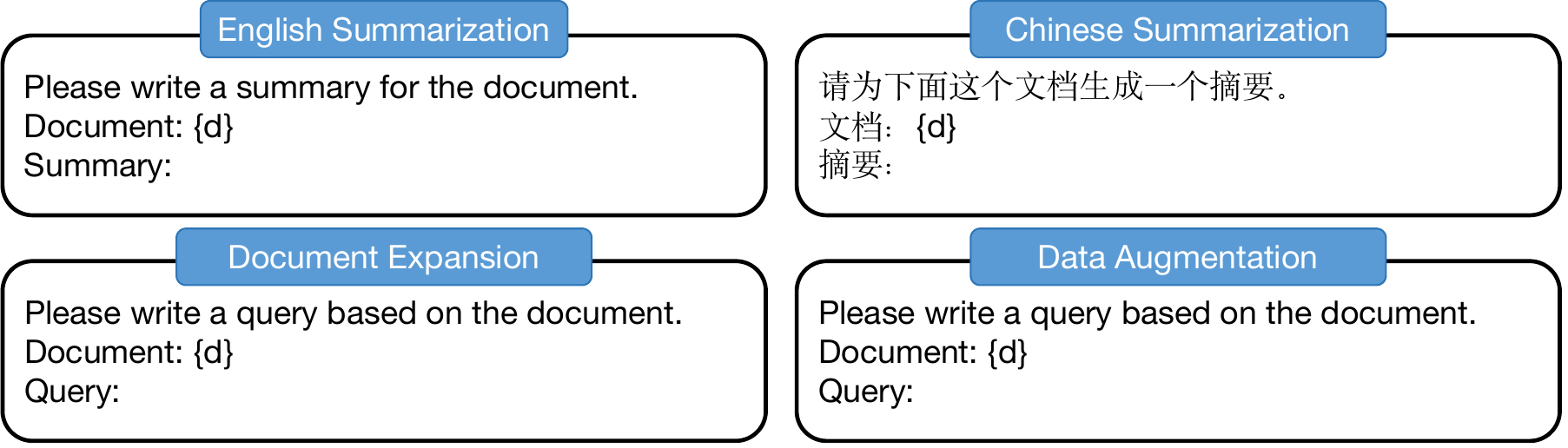}
    \caption{The templates used in our RLCF framework.}
    \label{fig:templates}
    \vspace{-5mm}
\end{figure}

More specifically, for each document $d\in\mathbb{G_d}$, we concatenate the pre-defined instruction (denoted as $\mathrm{Inst}$) as the input prefix for $d$, which can be defined as
\begin{equation}
    \label{eq:llm}
    o_d = \mathrm{LLM}\left(\mathrm{Inst}\oplus d\right),
\end{equation}
where $\oplus$ represents the concatenation operation.
The instruction templates are presented in Figure~\ref{fig:templates}.


\subsection{Model Optimization}
\label{sec:rlcf}

\noindent\textbf{Contrastive Feedback.}
To construct contrastive feedback with a group of similar documents, we first use an unsupervised dual encoder, as the representation encoder, to maps the responses and documents into a sematic space, and then compute the similarity between $o_{d_i}$ and each document $d_j\in \mathbb{G}=\left\{d_i\cup\mathbb{G}_{d_i}\right\}$. 
Intuitively, the similarity score between $d_i$ and $o_{d_i}$ would be the highest if LLM captures the key information of $d_i$ that distinguish it from other documents in $\mathbb{G}_{d_i}$.
Based on this intuition, we leverage a group-wise reciprocal rank (GRR) to implement contrastive feedback, which can be defined as
\begin{equation}
    \label{eq:batched-rr}
\textrm{GRR}({o_{d_i}}, \mathbb{G}) = \frac{\lambda}{\sum_{j=1}^{|\mathbb{G}|}\mathbb{I}(S(o_{d_i}, d_i)\leq S(o_{d_i}, d_j))},
\end{equation}
where $\mathbb{I}$ is an indicator function and $\lambda$ is a hyper-parameter.
$S(o_{d_i}, d_j)$ represents the similarity score of response $o_{d_i}$ and document $d_j$.
The similarity score is defined in Equation~\ref{eq:similarity}.

The computation of contrastive feedback is efficient. 
Although it necessitates $|\mathbb{G}|^2$ comparisons for the construction of a group-wise feedback, each document in a similar document group $\mathbb{G}$ requires encoding only once. 
These encoded document representations can then be reused to conduct inner product operations with multiple responses from $\mathbb{G}$.
For RLHF~\cite{ouyang2022training} and RLAIF~\cite{lee2023rlaif}, with the increase in the size of $\mathbb{G}$, their computational overhead  of constructing a group-wise contrastive feedback increase dramatically.
Each comparison involves the evaluation of a document-response pair, leading to quadratic computational complexity. 
Moreover, the expense associated with utilizing either an LLM annotator or a human annotator is considerably higher than that of dense retrieval.

Notably, GRR($o_{d_i}, \mathbb{G}$) is computed in a group-wise manner, specifically relying on the group $\mathbb{G}$.
Previous alignment techniques utilize a point-wise manner for computing reward scores, relying solely on $o_{d_i}$ and $d_i$.
This is problematic for adapting LLMs for IR because the distinctiveness of responses can only be measured in a group-wise manner.
In other words, we cannot determine the distinctiveness of a response for a document unless it is compared with responses generated for other documents.

\noindent\textbf{Optimization.}
Our objective is to optimize the policy model, i.e., the LLM, using contrastive feedback to generate responses that are desired in the context of IR.
We achieve this through reinforcement learning, specifically with the PPO~\cite{schulman2017proximal} algorithm.
We consider the GRR as the reward score for the entire response, and maximize it using the PPO algorithm.
Following prior study~\cite{ouyang2022training}, we also incorporate a term in the reward that penalizes the KL divergence between the optimized RL policy $\pi_{\phi}^{RL}$ with parameters $\phi$ and the original LLM $\pi$. 
The penalty term prevents the policy model from producing responses that diverge significantly from the vanilla LLM, thereby preserving the language capabilities of the policy model.
The full reward $\mathcal{R}$ could be written as
\begin{equation}
\label{eq:ppo_reward}
\mathcal{R}(d, o_d, \mathbb{G})=\textrm{GRR}(o_d, \mathbb{G})-\beta \log \left[\pi_\phi^{\mathrm{RL}}(o_d \mid d) / \pi(o_d \mid d)\right],
\end{equation}
where $\beta$ is a hyper parameter that balances the GRR and penalty term.

\section{Experimental Setup}
\subsection{LLM Applications in IR}
\label{sec:dowstream}
There are three popular applications of LLMs in IR, including document summarization \cite{Reid_2023, liu2023evaluating, huang2023citation, kou2023automated}, document expansion for sparse retrieval~\cite{gospodinov2023doc2query,wang2023query2doc} and data augmentation for dense retrieval~\cite{dai2022promptagator, bonifacio2022inpars}. We evaluate RLCF on the optimizing LLMs for these tasks. In the following, we briefly introduce why distinctiveness LLM outputs are crucial for these tasks.

\noindent\textbf{Document Summarization.}
Document summarization is a direct application of LLMs in many information systems, which is critical for reducing users' cognitive burden. Obviously, the summarization performance highly relies on the distinctive information captured by the summarization model (i.e., LLMs). However, due to a notable lake of distinctiveness in the responses of off-the-shelf LLMs, they often provide indistinguishable summaries for similar documents, such as the example shown in Figure~\ref{fig:llminir}. 

\begin{figure}
    \centering
    \includegraphics[width=\linewidth]{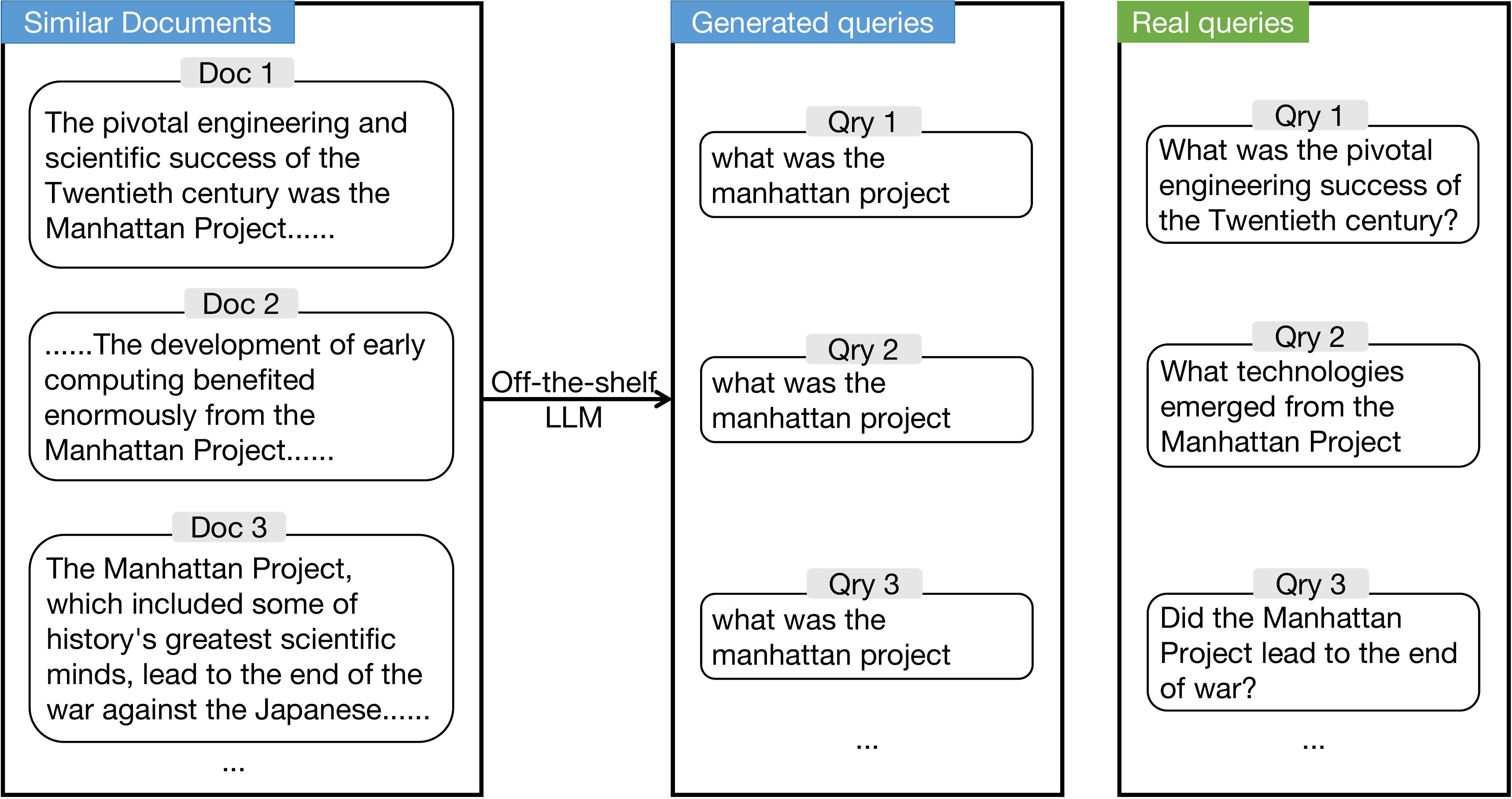}
    \caption{Illustrations of document expansion.}
    \label{fig:docexp}
    \vspace{-5mm}
\end{figure}

\noindent\textbf{Document Expansion for Sparse Retrieval.}
Document expansion is an effective technique for enhancing the performance of sparse retrieval~\cite{nogueira2019doc2query, nogueira2019document, gospodinov2023doc2query}, usually via mitigating the vocabulary mismatch between query and document.
To facilitate the accuracy of retrieval, the expansion of a document is desirable to be distinctive compared with other similar documents, for which we find out that vanilla LLMs are not well-aligned. Figure~\ref{fig:docexp} shows an example of FLAN-T5 generating three identical queries for similar documents, which could potentially undermine the performance of downstream sparse retrieval. 

\noindent\textbf{Data Augmentation for Dense Retrieval.}
For dense retrieval, data augmentation using LLMs is an effective solution for handling out-of-distribution (OOD) documents (e.g., newly-created contents) and scenarios~\cite{dai2022promptagator}, where the labeled data is usually scarce.
In particular, existing methods propose to generate synthetic queries for the OOD documents, and train a retrieval model on such synthetic query-document pairs with contrastive learning~\cite{chen2020simple}. Notably, improving the distinctiveness of the generated query can help a retrieval model to develop more robust capability of relevance matching. 
In contrast, if a generated query is trivial, vague, or similar to other generated queries, such data can hardly be used to train an effective retrieval model.

\subsection{Datasets}
\noindent\textbf{Document Summarization.}
To compare the effectiveness for different alignment methods on document summarization, we perform experiments on two datasets: LCSTS for Chinese and Gigaword for English.
\textbf{LCSTS}~\cite{hu2015lcsts} is a widely used dataset employed for Chinese text summarization task. It was created to facilitate research and development in the field of short text summarization. 
\textbf{Gigaword}~\cite{see2017get} is extensively utilized in English text summarization research, comprises substantial news articles and their associated headline summaries. 
This dataset is known for its extensive coverage of diverse topics and its massive size, which makes it a valuable resource for training and evaluating text summarization models.
The corpus of LSCTS and Gigaword contain millions of documents.
The corpus of other datasets, such as CNN/Daily Mail~\cite{nallapati2016abstractive}, WikiSum~\cite{Bert2BertWikiSummaryPersian} and PubMed~\cite{cohan2018discourse}, are typically of smaller size. 
Documents in these datasets diverge significantly, which makes them easy to be differentiated and not appropriate to simulate real IR scenarios. Therefore, these datasets are not the primary focus of this paper.

\noindent\textbf{Document Expansion and Data Augmentation for Document Retrieval.}
We use the \textbf{BEIR}~\cite{thakur2021beir} to evaluate the effectiveness of data augmentation. BEIR~\cite{thakur2021beir} is a widely-used benchmark for document retrieval.
It comprises a variety of tasks, including passage retrieval, entity retrieval, fact checking, and others. The dataset includes a broad range of domains, such as medical, finance, and science. 
NQ~\cite{kwiatkowski2019natural} is not included in our experiments, because it is presented in the training stage of FLAN-T5~\cite{dai2022promptagator}.

\subsection{Implementation Details}
We only utilize the corpus of summarization datasets (i.e., LCSTS and Gigaword) to conduct RLCF optimization and other model-based LLM alignment methods.
Subsequently, inference is directly performed for the optimized LLM, followed by comparative analysis.
In our experiments, we employ FLAN-T5~\cite{chung2022scaling} as the backbone of LLMs for English datasets, which is an encoder-decoder architecture. 
We perform experiments using FLAN-T5 models with 770M, 3B, and 11B parameters, respectively.
For the Chinese dataset, we utilize BELLE-7B-2M~\cite{BELLE}, 
which is a decoder-only architecture and achieve promising instruction-following ability in Chinese.
Although GPT-3.5 and GPT-4 demonstrate superior performance, the undisclosed parameters hinder the training and evaluation of alignment. We use Contriever~\cite{izacard2021unsupervised} as the unsupervised dual-encoder $M$ in RLCF. 
Each document is limited to the first 512 tokens, and any tokens beyond 512 are truncated.
The $\lambda$ in Equation~\ref{eq:batched-rr} is 10, and the group size used in our experiments is 32, i.e., $|\mathbb{G}|$ is 32.
The sparse retrieval method used in document expansion experiments is BM25~\cite{robertson2009probabilistic}, implemented by Anserini~\footnote{https://github.com/castorini/anserini} with default parameters.

To maximize the efficient utilization of GPU memory, we optimize all the parameters in FLAN-T5 with 770M parameters, the last 23 layers of FLAN-T5 with 3B parameters, the last 4 layers of FLAN-T5 with 11B parameters, and the last 12 layers of BELLE-7B-2M, respectively.
During text generation, 
we simply use the greedy decoding strategy.

All experiments are implemented with PyTorch and Huggingface. 
DeepSpeed with ZeRO stage 2 is utilized for efficient training.
All the training and evaluation are conducted on 8 NVIDIA Tesla A100 GPUs (with 40G RAM).

\subsection{Evaluation}
\label{sec:humansetting}
\noindent\textbf{Automatic Evaluation.}
For data augmentation of dense retrieval, we directly utilize the traditional metrics of document retrieval, i.e., Mean Reciprocal Rank (MRR), Recall, and Normalized Discounted Cumulative Gain (NDCG).
We introduce Rouge-diff as an evaluation metric for document summarization, aimed at assessing the distinctiveness of summaries within similar documents.
The Rouge-diff is a variant of Rouge-N, which is defined as
\begin{equation}
    \label{eq:rougediff}
    \textrm{Rouge-diff}_{o_{d_i}} = \frac{|set(o_{d_i})\cap (set(d_i)\setminus set(\cup\mathbb{G}_{d_i}))|}{|set(d_i)\setminus set(\cup\mathbb{G}_{d_i})|}.
\end{equation}
Here, $set(t)$ represents the tokens of text $t$ after deduplication, and $|set(t)|$ denotes the number of tokens in $set(t)$.
Additionally, we report GRR for summarization evaluation, which is defined in Equation (\ref{eq:batched-rr}).

\noindent\textbf{Human Evaluation.}
To make the evaluation more convincing, we further conduct human evaluation on summaries.
Firstly, we randomly sample 200 documents that are not used in the RLCF optimization.
Subsequently, we identify the 3 most similar documents for each of these documents, forming 200 groups documents with 4 documents in each group. 
Finally, we generate summaries for these 200*4 documents using both vanilla LLMs and RLCF-optimized LLMs, as well as GPT-4.
We recruit three annotators from a pool of Ph.D. students, each with expertise in areas such as natural language processing and information retrieval.
We provide annotation guidelines to our human experts and instruct them to conduct a three-level annotation. 
The annotation guidelines involve three dimensions: distinctiveness, correctness and concision. 
\textit{Distinctiveness} refers to the ability of the summary to distinguish itself from similar documents. It requires the summary to highlight unique and critical points that set it apart from other similar documents.
\textit{Correctness} represents the accuracy and completeness of the information presented. 
\textit{Concision} concerns the brevity of the summary. A concise summary effectively conveys the main points of the original document in as few words as possible.
The annotation process in RLCF is conducted at the group level, wherein the ultimate decision regarding superior responses is made through comprehensive evaluation.
We conduct comparisons between the vanilla LLM and the RLCF-optimized LLM, as well as between the RLCF-optimized LLM and GPT-4.	
During the process of annotation, two LLMs are randomly designated as LLM A and LLM B to avoid bias.
The annotator’s task is to assess the quality of the summaries and determine which one is superior. 
The decision is based on the above aspects of the summaries, i.e., distinctiveness, correctness, and concision.


\section{Experimental Results}
\label{sec:experiments}
\subsection{Document Summarization}
\label{sec:sum}
For Chinese document summarization, we employ BELLE-7B-2M, and for English document summarization, we utilize FLANT5-3B as the initial parameters of LLMs.
We conduct both automatic evaluation and human evaluation for document summarization.

\noindent\textbf{Automatic Evaluation.} 
We randomly select 512 documents that are not used in the RLCF optimization to form the initial test set.
Subsequently, the four documents most similar to each document in the initial test set are retrieved by a dual-encoder, thereby extending the initial test set and making the evaluation challenging. 
As a result, the final test set consists of 2048 documents.

The experimental results are presented in Table~\ref{table:summarization}.
From this table, we can draw the following findings:
\begin{itemize}
    \item RLCF optimization significantly improves the Rouge-diff on the test set, demonstrating its effectiveness on document summarization in IR context.
    \item RLCF optimization leads to significant improvements on both Chinese and English datasets, highlighting its effectiveness across different languages as well as various parameter scales and architectures.
    \item GPT-3.5 and GPT-4.0 demonstrate superior performance in generating distinctive responses compared to publicly available LLMs. However, as the parameters of GPT-3.5 and GPT-4.0 have not been released, conducting RLCF optimization experiments on them is currently unfeasible.
    \item Model-based alignment methods (i.e., RLAIF and RLCD) cannot outperform GPT-3.5 and GPT-4.0 on LCSTS, owing to the inherent linguistic capabilities of the Chinese LLM BELLE-7B-2M, which are comparatively weaker than those of the English LLM FLAN-T5. The unreliable feedback constrains the effectiveness of model-based alignment methods.
    \item In comparison to point-wise based alignment methods, RLCF significantly outperforms them, underscoring the effectiveness of the group-wise manner.
\end{itemize}

\begin{figure}
    \centering
    \subfigure[LCSTS]{
        \includegraphics[width=0.41\linewidth]{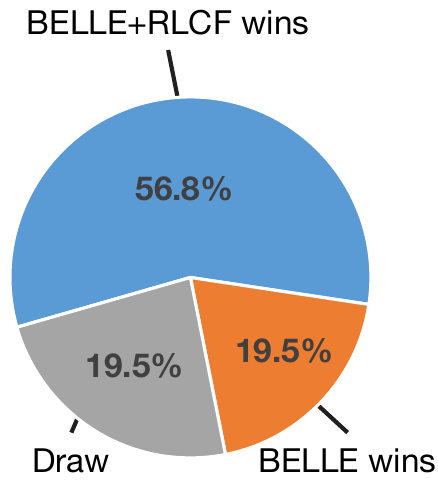}
    }\hspace{2mm}
    \subfigure[Gigaword]{
        \includegraphics[width=0.41\linewidth]{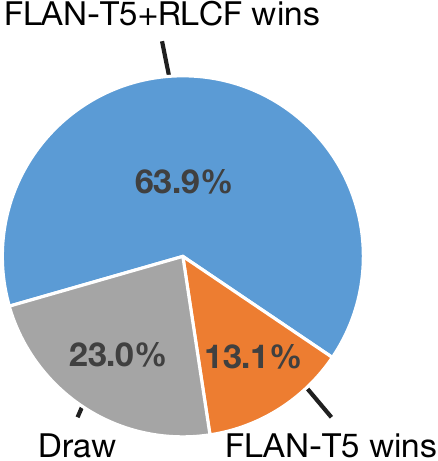}
    }
    \subfigure[LCSTS]{
        \includegraphics[width=0.37\linewidth]{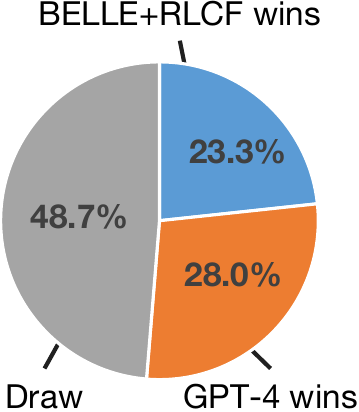}
    }\hspace{2mm}
    \subfigure[Gigaword]{
        \includegraphics[width=0.39\linewidth]{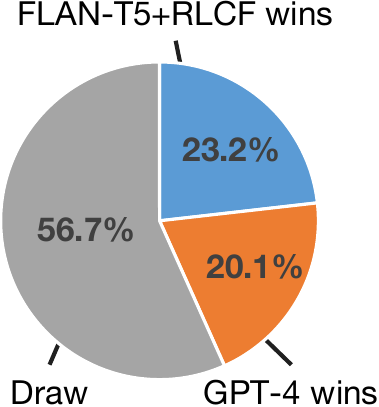}
    }
    \caption{The results of human evaluation.}
    \label{fig:human}
\end{figure}

\noindent\textbf{Human Evaluation.}
We also incorporate human evaluation in our experiments.
The settings of human evaluation are presented in Section~\ref{sec:humansetting}.
The evaluation results are presented in Figure~\ref{fig:human}. 
From this figure, we can draw the following conclusions:
\begin{itemize}
    \item Responses generated by the RLCF-optimized LLMs contain more distinctive information than those produced by vanilla LLMs, making them more suitable for IR scenarios.
    \item Gigaword's results are superior to those of LCSTS. This discrepancy can be attributed to Gigaword's larger corpus size (3.8 million vs. 2.4 million) and the higher degree of similarity among its documents, making it has more similar documents to construct contrastive feedback.
    \item The RLCF-optimized LLMs exhibit performance on par with GPT-4 on both Chinese and English datasets. Given the massive scale of parameters in GPT-4, the effectiveness of RLCF is remarkable, as the LLMs used in our experiments have only 3-7 billion parameters.
    \item GPT-4 slightly outperforms the RLCF-optimized LLM on the Chinese dataset. This discrepancy could be attributed to the inherent disparities in the fundamental capabilities of LLMs. The publicly available English LLM is superior to its Chinese counterpart.
\end{itemize}


\begin{table}[]
\caption{Experimental results of document summarization on LCSTS and Gigaword. Significant improvement or degradation w.r.t. vanilla LLM is indicated (+/-) (\emph{p-value}$\leq$0.05).}
\label{table:summarization}
\begin{tabular}{l|cc|cc}
\hline\hline
            \multirow{2}{*}{Model(\#params)}               & \multicolumn{2}{c|}{LCSTS}                  & \multicolumn{2}{c}{Gigaword}                \\ \cline{2-5} 
                           & \multicolumn{1}{c|}{Rouge-diff} & GRR & \multicolumn{1}{c|}{Rouge-diff} & GRR \\ \hline
GPT-3.5(20B)                        & \multicolumn{1}{c|}{23.8}       & 90.9     & \multicolumn{1}{c|}{15.5}       & {78.9}      \\ \hline
GPT-4.0(1.8T)                         & \multicolumn{1}{c|}{25.6}       & 90.9     & \multicolumn{1}{c|}{{17.6}}       & {78.9}    \\ \hline\hline
\begin{tabular}[c]{@{}l@{}}Vanilla LLM\\ \footnotesize BELLE(7B)/FLAN-T5(3B)\end{tabular}                       & \multicolumn{1}{c|}{22.1}       & 90.4      & \multicolumn{1}{c|}{11.9}       & 75.2      \\ \hline
\multicolumn{1}{r|}{w/ RLAIF} & \multicolumn{1}{c|}{22.7}       &    90.6    & \multicolumn{1}{c|}{18.5}       & {77.3}      \\ \hline
\multicolumn{1}{r|}{w/ RLCD} & \multicolumn{1}{c|}{\underline{23.4}}       &     \underline{90.8}   & \multicolumn{1}{c|}{\underline{19.3}}      & \underline{78.2}      \\ \hline
\multicolumn{1}{r|}{w/ RLCF} & \multicolumn{1}{c|}{\textbf{32.2}$^+$}       & \textbf{91.7}$^+$      & \multicolumn{1}{c|}{\textbf{32.5}$^+$}       & \textbf{80.9}$^+$      \\
\hline\hline
\end{tabular}
\vspace{-5mm}
\end{table}

\begin{table*}[]
\caption{Experimental results of sparse retrieval methods on BEIR. 
Significant improvement or degradation w.r.t. DADR is indicated (+) (\emph{p-value}$\leq$0.05). DocExp is the abbreviation for document expansion. The metrics used in this table is NDCG@10.}
\label{table:docexp}
\begin{tabular}{lccccccc}
\hline\hline
\multicolumn{1}{c|}{}                          & \multicolumn{1}{l|}{}                      & \multicolumn{1}{c|}{}                       & \multicolumn{5}{c}{DocExp}                                                                                                                              \\ \cline{4-8} 
\multicolumn{1}{l|}{\multirow{-2}{*}{Dataset}} & \multicolumn{1}{l|}{\multirow{-2}{*}{ISL}} & \multicolumn{1}{c|}{\multirow{-2}{*}{BM25}} & \multicolumn{1}{c|}{FLAN-T5(3B)} & \multicolumn{1}{c|}{w/ RLAIF}   & \multicolumn{1}{c|}{w/ RLCD}    & w/ RLCF                              & Improv. \\ \hline
\multicolumn{8}{c}{\#Avg. Doc. Length\textless{}200}                                                                                                                                                                                                                                                \\ \hline
\multicolumn{1}{l|}{FiQA}                      & \multicolumn{1}{c|}{1.27}                  & \multicolumn{1}{c|}{.162}                   & \multicolumn{1}{c|}{{\ul .215}}    & \multicolumn{1}{c|}{.208}       & \multicolumn{1}{c|}{.207}       & \textbf{.216}                        & 0.5\%   \\ \hline
\multicolumn{1}{l|}{SCIDOCS}                   & \multicolumn{1}{c|}{1.29}                  & \multicolumn{1}{c|}{.133}                   & \multicolumn{1}{c|}{{\ul .138}}    & \multicolumn{1}{c|}{{\ul .138}} & \multicolumn{1}{c|}{.137}       & {\color[HTML]{333333} \textbf{.139}} & 0.7\%   \\ \hline
\multicolumn{1}{l|}{ArguAna}                   & \multicolumn{1}{c|}{1.32}                  & \multicolumn{1}{c|}{.231}                   & \multicolumn{1}{c|}{.253}          & \multicolumn{1}{c|}{{\ul .259}} & \multicolumn{1}{c|}{.253}       & {\color[HTML]{333333} \textbf{.270}$^{+}$} & 4.2\%   \\ \hline
\multicolumn{1}{l|}{TREC-COVID}                & \multicolumn{1}{c|}{1.49}                  & \multicolumn{1}{c|}{.456}                   & \multicolumn{1}{c|}{{\ul .563}}    & \multicolumn{1}{c|}{.560}       & \multicolumn{1}{c|}{.562}       & {\color[HTML]{333333} \textbf{.582}$^{+}$} & 3.4\%   \\ \hline
\multicolumn{1}{l|}{Avg.} & \multicolumn{1}{c|}{1.34} & \multicolumn{1}{c|}{.246}                   & \multicolumn{1}{c|}{{\ul .292}}    & \multicolumn{1}{c|}{.291}       & \multicolumn{1}{c|}{.290}       & {\color[HTML]{333333} \textbf{.302}} & 3.4\%   \\ \hline
\multicolumn{8}{c}{\#Avg. Doc. Length\textgreater{}200}                                                                                                                                                                                                                                             \\ \hline
\multicolumn{1}{l|}{SciFact}                   & \multicolumn{1}{c|}{1.21}                  & \multicolumn{1}{c|}{.666}                   & \multicolumn{1}{c|}{.668}          & \multicolumn{1}{c|}{{\ul .671}} & \multicolumn{1}{c|}{.669}       & \textbf{.674}$^{+}$                        & 0.4\%   \\ \hline
\multicolumn{1}{l|}{NFCorpus}                  & \multicolumn{1}{c|}{1.26}                  & \multicolumn{1}{c|}{.318}                   & \multicolumn{1}{c|}{\textbf{.325}} & \multicolumn{1}{c|}{.318}       & \multicolumn{1}{c|}{.319}       & {\ul .320}                           & -       \\ \hline
\multicolumn{1}{l|}{Touche}                    & \multicolumn{1}{c|}{1.39}                  & \multicolumn{1}{c|}{.489}                   & \multicolumn{1}{c|}{.489}          & \multicolumn{1}{c|}{.489}       & \multicolumn{1}{c|}{{\ul .494}} & {\color[HTML]{333333} \textbf{.498}$^{+}$} & 0.8\%   \\ \hline
\multicolumn{1}{l|}{Avg.} & \multicolumn{1}{c|}{1.28} & \multicolumn{1}{c|}{.491}                   & \multicolumn{1}{c|}{{\ul .494}}    & \multicolumn{1}{c|}{.493}       & \multicolumn{1}{c|}{{\ul .494}}       & {\color[HTML]{333333} \textbf{.497}} & 0.6\%   \\ \hline\hline
\end{tabular}
\end{table*}

\subsection{Document Expansion for Sparse Retrieval}
Document expansion is an effective technique to alleviate the vocabulary mismatch, thus improving the performance of sparse retrieval.
We generate five queries per document for the purpose of document expansion, and employ BM25~\cite{robertson2009probabilistic} to assess the effectiveness of the expanded contents.
Due to constraints in computational resources, we restricted our document expansion experiments to seven datasets from BEIR~\cite{thakur2021beir}, each containing fewer than one million documents.
The generation of five queries per document for a dataset comprising millions of documents requires a significant investment of GPU hours, amounting to thousands.

The effectiveness of document expansion is influenced by inherent characteristics of the dataset.
Therefore, we incorporate statistical information, including the average length of documents and intra-list similarity, into Table~\ref{table:docexp}.
We divided this table into two groups based on the average document length to facilitate the analysis.
Intra-list similarity (ILS)~\cite{ziegler2005improving, xie2023t2ranking} is defined as
\begin{equation}
    \mathbf{I L S}_{\mathcal{D}}=\frac{\sum_{i=1}^{|\mathcal{D}|} \sum_{j=i+1}^{|\mathcal{D}|} S({d}_i, {d}_j)}{\sum_{i=1}^{|\mathcal{D}|} \sum_{j=i+1}^{|\mathcal{D}|} 1},
\end{equation}
where $S(d_i, d_j)$ is the similarity score between $d_i$ and $d_j$, defined in Equation~\ref{eq:similarity}.

The experimental results are shown in Table~\ref{table:docexp}.
Compared with document expansion from off-the-shelf LLM, the average BM25 performance is improved from .246 to .292 when the average document length is less than 200.
This highlights that the problem of mismatched vocabulary is more prominent in shorter documents. 
Thus, shorter documents are better suited for assessing the effectiveness of various alignment techniques in document expansion.
Besides, DocExp w/ RLCF outperforms other methods on six out of seven datasets, with notable excellence on ArguAna and TREC-COVID. 
These two datasets exhibit higher ILS scores. 
A heightened ILS score signifies increased similarity among documents in the corpus, indicating a more pronounced inclination for distinctive expanded contents. 
This demonstrates that as the similarity among documents within the corpus grows, LLMs increasingly depend on RLCF to enhance their performance on generating distinctive responses.


\begin{table*}[]
\caption{Experimental results of dense retrieval methods on BEIR.
Significant improvement or degradation w.r.t. DADR is indicated (+) (\emph{p-value}$\leq$0.05). DataAug is the abbreviation of data augmentation.}
\label{table:beir}
\begin{tabular}{r|c|c|c|c|c|c|c|c|c|c|c|c}\hline\hline
Dataset($\rightarrow$)      & \small ArguAna & \small FiQA & \small NFCorpus & \small SCIDOCS & \small SciFact & \small COVID	 & \small Touche & \small DBPedia & \small HotpotQA& \small Fever & \small MARCO &  \small Avg.\\\hline
\multicolumn{1}{l|}{Method($\downarrow$)}                    & \multicolumn{10}{c}{NDCG@10}                                                                                                                                                                                                                                                                                                                                                                                                                                                                  \\ \hline
\multicolumn{1}{l|}{\begin{tabular}[l]{@{}l@{}}DataAug\\ \footnotesize FLAN-T5(3B)\end{tabular}}  & \underline{.166}    & {.152} & .144 & \underline{.073} & .339 & {.452} & \underline{.155} & {.173} & {.335}& .292&\underline{.133}& {.219} \\\hline
w/ RLAIF                          & {.162} & \underline{.159} & .144 & .072 & {.345} & .467 & .145 & \underline{.183} &\underline{.348}& .291 & .128 & \underline{.222} \\\hline
w/ RLCD                           & .125 & {.144} & \underline{.153} & .065 & \underline{.352} & \underline{.473} & .112 & {.166} & .340& \underline{.314} &.122& .215 \\\hline
w/ RLCF                           & \textbf{.210}$^{+}$ & \textbf{.168}$^{+}$ & \textbf{.155} & \textbf{.081}$^{+}$ & \textbf{.379}$^{+}$ & \textbf{.502}$^{+}$ & \textbf{.164}$^{+}$ & \textbf{.192}$^{+}$ & \textbf{.390}$^{+}$& \textbf{.347}$^{+}$ & \textbf{.144}$^{+}$ & \textbf{.245}\\\cline{2-13}
Improv.& 26.5\% & 5.7\% & 1.3\%  & 11.0\% & 7.7\% & 6.1\% & 5.8\% & 4.9\% & 12.1\% & 10.5\% & 8.3\% & 10.4\% \\\hline
                     & \multicolumn{10}{c}{Recall@100}                                                                                                                                                                                                                                                                                                                                                                                                                                                                  \\ \hline
\multicolumn{1}{l|}{\begin{tabular}[l]{@{}l@{}}DataAug\\ \footnotesize FLAN-T5(3B)\end{tabular}} & \multicolumn{1}{c|}{\underline{.795}}        & \multicolumn{1}{c|}{.444} & \multicolumn{1}{c|}{.184}         & \multicolumn{1}{c|}{.203}        & \multicolumn{1}{c|}{.708}        & \multicolumn{1}{c|}{.068}       & \multicolumn{1}{c|}{.279}   & \multicolumn{1}{c|}{.280}           & \multicolumn{1}{c|}{.526}  &.574& \underline{.814} &   .442   \\ \hline
\multicolumn{1}{r|}{w/ RLAIF}      & \multicolumn{1}{c|}{.789}        & \multicolumn{1}{c|}{\underline{.448}} & \multicolumn{1}{c|}{\underline{.197}}         & \multicolumn{1}{c|}{\underline{.207}}        & \multicolumn{1}{c|}{\underline{.737}}        & \multicolumn{1}{c|}{{.071}}       & \multicolumn{1}{c|}{\underline{.281}}   & \multicolumn{1}{c|}{.298}           & \multicolumn{1}{c|}{\underline{.527}} & .579 &  .809  &  \underline{.445}    \\ \hline
\multicolumn{1}{r|}{w/ RLCD}       & \multicolumn{1}{c|}{.724}        & \multicolumn{1}{c|}{.427} & \multicolumn{1}{c|}{.191}         & \multicolumn{1}{c|}{.185}        & \multicolumn{1}{c|}{.717}        & \multicolumn{1}{c|}{\underline{.072}}       & \multicolumn{1}{c|}{.238}   & \multicolumn{1}{c|}{\underline{.299}}           & \multicolumn{1}{c|}{.525} & \underline{.601} & .792 &  .434    \\ \hline
\multicolumn{1}{r|}{w/ RLCF}       & \multicolumn{1}{c|}{\textbf{.844}$^{+}$}        & \multicolumn{1}{c|}{\textbf{.450}} & \multicolumn{1}{c|}{\textbf{.202}$^{+}$}         & \multicolumn{1}{c|}{\textbf{.209}}        & \multicolumn{1}{c|}{\textbf{.744}$^{+}$}        & \multicolumn{1}{c|}{\textbf{.076}$^{+}$}       & \multicolumn{1}{c|}{\textbf{.304}$^{+}$}   & \multicolumn{1}{c|}{\textbf{.319}$^{+}$}           & \multicolumn{1}{c|}{\textbf{.564}$^{+}$}    & \textbf{.646}$^{+}$ &  \textbf{.834}$^{+}$  & \textbf{.471} \\ \cline{2-13}
Improv.& 6.2\% & 0.4\% & 2.5\%  & 1.0\% & 0.9\% & 5.6\% & 9.2\% & 6.7\% & 7.0\% &7.5\% & 2.5\% &5.8\% \\\hline\hline
\end{tabular}
\end{table*}

\subsection{Data Augmentation for Dense Retrieval}
\label{sec:fsdr}


The experimental results of BEIR are presented in Table~\ref{table:beir}.
Notably, the settings used in this table are entirely zero-shot. 
To facilitate the evaluation of data augmentation's quality, we utilize the BERT-based-uncased~\cite{devlin2018bert} as the initial parameters of dense retrieval.
From this table, we can draw the following findings:
\begin{itemize}
    \item Across all datasets, RLCF-optimized LLMs consistently outperform other alignment methods in data augmentation for dense retrieval.
    This demonstrates the effectiveness of our RLCF framework for aligning the capability of LLMs with the data augmentation in IR.
    \item Compared with DataAug without any alignment, point-wise alignment methods, RLAIF and RLCD, do not exhibit significant improvements. This underscores the superiority of group-wise contrastive feedback used in RLCF.
\end{itemize}

\begin{figure}
    \centering
    \subfigure[MRR@10]{
        \includegraphics[width=0.45\linewidth]{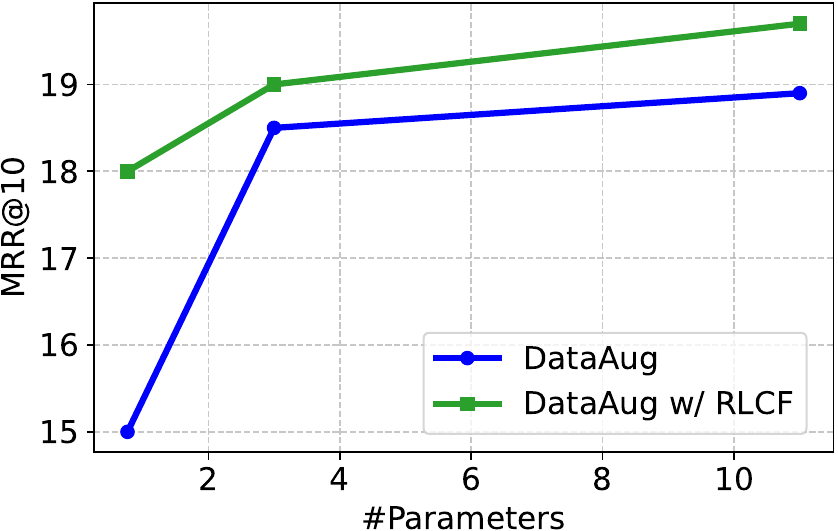}
    }\hspace{2mm}
    \subfigure[Recall@1000]{
        \includegraphics[width=0.45\linewidth]{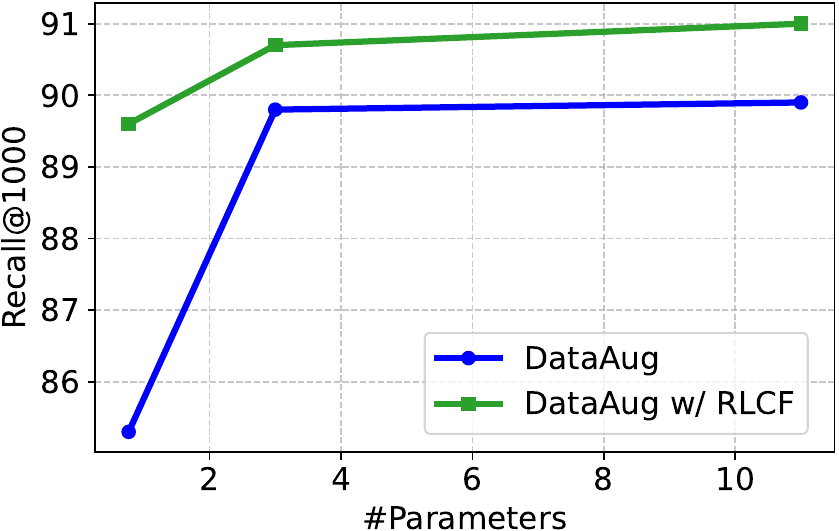}
    }
    \caption{Scaling law of LLMs on the application of data augmentation. The x-axis is the number of parameters in billions.}
    \label{fig:scaling_law}
    \vspace{-5mm}
\end{figure}

To further analyze the impact of different scales of LLMs on DADR, we conduct an analysis of scaling LLMs on MS-MARCO. We use LLMs with 770 million, 3 billion, and 11 billion parameters in the experiment. 
The results are illustrated in Figure~\ref{fig:scaling_law}, which demonstrates that RLCF can consistently improve the performance of DADR with different parameter scales.
Notably, the metrics we employ are MRR@10 and Recall@1000, as these are widely used metrics in the MS-MARCO benchmark.
As LLMs are trained under a generic domain for query generation, we notice that when the language model has a small number of parameters (i.e., 770M), substantial generated queries be like "What is the main idea of this document?". 
Additionally, as depicted in Figure~\ref{fig:scaling_law}, the RLCF-optimized LLMs adhere to the scaling law and outperform the LLMs with equivalent parameters.

\subsection{Efficiency Analysis}

\begin{figure}
    \centering
    \includegraphics[width=0.7\linewidth]{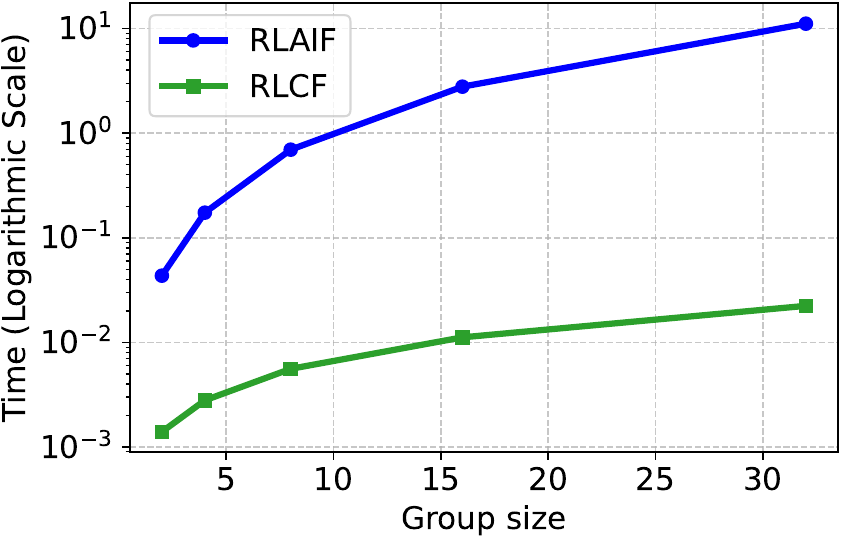}
    \caption{The relationship between the inference time and group size for RLAIF and RLCF. The y-axis represents the logarithmic scale of time (seconds) for better visualization.}
    \label{fig:Inference}
    \vspace{-5mm}
\end{figure}

In this section, we conduct an analysis of the group-wise feedback computation inference times.
As depicted in Figure~\ref{fig:Inference}, the inference time of RLAIF exhibits an approximate exponential increase with the growth of the group size, due to the unavailability of reusable document representations in feedback computation.
The time overhead in RLCF is primarily lies in encoding documents and responses, while the computational cost of inner product calculations is negligible. 
Therefore, RLCF displays an approximately linear correlation with group size.
Notably, the LLM used in Figure~\ref{fig:Inference} for RLAIF is FLAN-T5-3B, while the dual encoder used in RLCF is Contriever with only 110 million parameters.
When the group size is 32, RLAIF requires a time overhead of 11.3 seconds, utilizing 6.2 GB of additional GPU memory for group-wise feedback computation. Upon increasing the parameters of the LLM to 7 billion, the time overhead escalates to 50.9 seconds, and additional GPU memory usage rises to 14.3 GB.
As a comparison, RLCF requires only an additional GPU memory overhead of 0.25GB and a time overhead of 2.2e-3 seconds.
Consequently, RLCF is an efficient alignment method for adopting LLMs to IR.

\subsection{Case Study}
\label{sec:case}
\begin{figure}
    \centering
    \includegraphics[width=\linewidth]{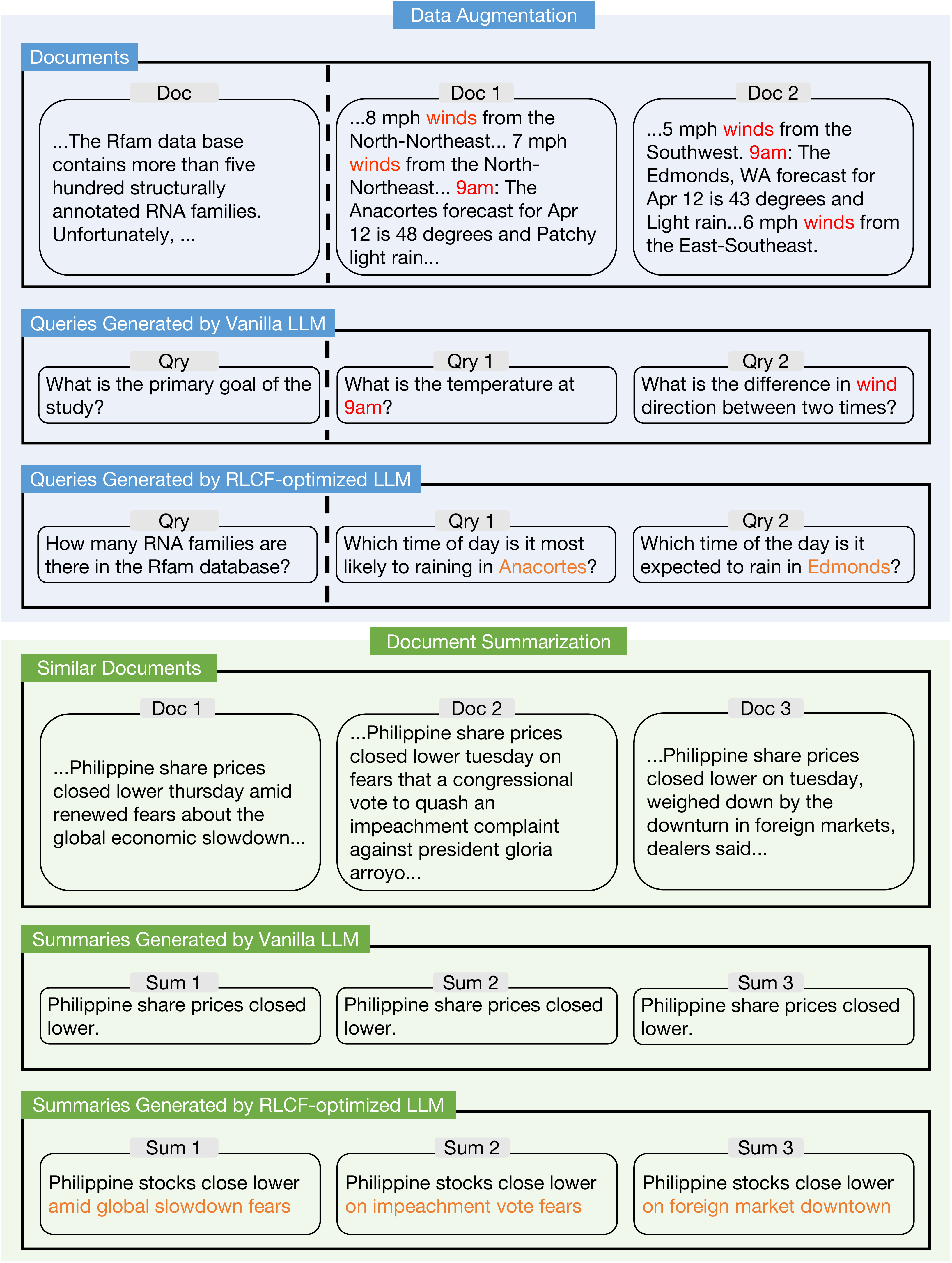}
    \caption{The cases of responses generated by vanilla LLMs and RLCF-optimized LLMs for highly similar documents.}
    \label{fig:casestudy}
    \vspace{-5mm}
\end{figure}

In this subsection, we present several cases to better illustrate the effectiveness of RLCF, as shown in Figure~\ref{fig:casestudy}.
Since LLMs are employed for generating queries in both document expansion and data augmentation, we represent these two applications solely by the use of "Data Augmentation" in Figure~\ref{fig:casestudy}.
For the task of data augmentation, it is evident that the queries generated by vanilla LLMs lack distinctiveness.
In the first case, the query generated by vanilla LLMs could even match all documents.
In the second case, despite the generated queries being relatively more relevant to the documents, they still lack distinctiveness. A query generated by vanilla LLMs for one document can still match another document. For example, the generated query, i.e., "What is the temperature at 9am?", can also be answered by another document. 
As the generated query and its corresponding document are positive training examples of each other in the contrastive learning training of dense retrieval, the lack of specific generated queries leads to the false negative problem, which hampers the performance of dense retrieval models~\cite{dai2022promptagator,ren2021rocketqav2}.
Unlike vanilla LLMs, RLCF-optimized LLMs can alleviate this problem and thus improve the effectiveness of data augmentation.

For the task of document summarization, we can see that the summaries generated by vanilla LLM are all the same for these similar documents. Despite that the generated summarizes are accurate for individual documents, they are not suitable within the pipeline of IR. In the context of IR, once a user submits a query, the search engine retrieves a collection of documents relevant to the query. These documents naturally possess a high degree of similarity. Generating differentiable summaries for such highly similar documents aids users in filtering and identifying their desired documents.
As shown in Figure~\ref{fig:casestudy}, after RLCF optimization, the summaries generated by the LLMs not only precisely summarize the main idea of the document, i.e., the lowering of Philippine stocks, but also provide specific reasons based on their corresponding documents.
The summaries generated by RLCF-optimized LLMs demonstrate a higher degree of distinctiveness towards their respective documents, making them more suitable for IR scenarios.

Therefore, through RLCF optimization, the capabilities of LLMs can be effectively aligned with the context of IR, resulting in the generation of more distinctive summaries and queries for documents.

\section{Conclusion}
In this work, we propose a novel framework, namely RLCF, that leverages contrastive feedback to optimize large language models through reinforcement learning.
The capabilities of LLMs could be aligned with the requirements of information retrieval through the proposed RLCF in an unsupervised manner.
Specifically, we construct groups of similar documents by an unsupervised dual encoder model, and then use an LLM to generate a response for each document within the group.
Next, we leverage the generated response to construct a contrastive feedback for the LLM optimization, which is implemented by a group-wise reward function, i.e., group-wise reciprocal rank. By doing this, the LLM can be optimized via PPO algorithm to be aligned with the requirements of information retrieval.
%
We conduct experiments on three popular applications of LLMs in information retrieval, demonstrating the effectiveness of our proposed RLCF.
The RLCF-optimized LLMs could generate document summaries with more distinctiveness, which could help users better distinguish relevant documents from similar yet less relevant documents in IR scenarios.
Also, the RLCF-optimized LLMs could generates distinctive queries for document expansion and data augmentation, and thus achieving promising performance.
To evaluate the effectiveness of summarization in the proposed setting, we introduce an automatic metric rouge-diff, a variant of rouge score, which is calculated in the group-wise manner. 
Besides, we also conduct human evaluation for the LLMs responses to evaluate the distinctiveness.
The experimental results demonstrate the superiority of RLCF in handling both Chinese and English languages across diverse parameter scales and model architectures.
In future work, we suggest exploring more domains which could use the RLCF for optimization, such as style transfer, harmless alignment, helpfulness alignment and etc.

\bibliographystyle{ACM-Reference-Format}
\bibliography{sample-base}
\appendix
\end{document}